\documentstyle[twoside,epsfig,floatflt]{hsproc}

\def\runauthor{Matthew Wing}
\def\shorttitle{Photon Structure and Hadronic Final States at ZEUS}

\begin{document}

\begin{flushright}
UCL/HEP 98-02
\end{flushright}

\title{Photon Structure and the Hadronic Final State in
Photoproduction Processes at ZEUS}
\author{Matthew Wing \email{wing@zow.desy.de}}{University College London, UK}

\abstract{Advances in knowledge of the structure of the photon and
tests of perturbative QCD have been made using the increased
luminosity with the ZEUS detector at HERA. Events with a low photon
virtuality  and two high transverse energy jets have been
studied. Measurements of inclusive dijet and  multijet production are
herein compared to Next-to-Leading-Order calculations. In order to
analyse the structure of the photon, dijet production of quasi-real
and virtual photons and dijet production containing $D^{*\pm}$ mesons
were measured.}

\vspace{-0.6cm}

\section{Introduction}

\begin{figure}[htp]
\vspace{-0.35cm}
\begin{center}
\epsfig{file=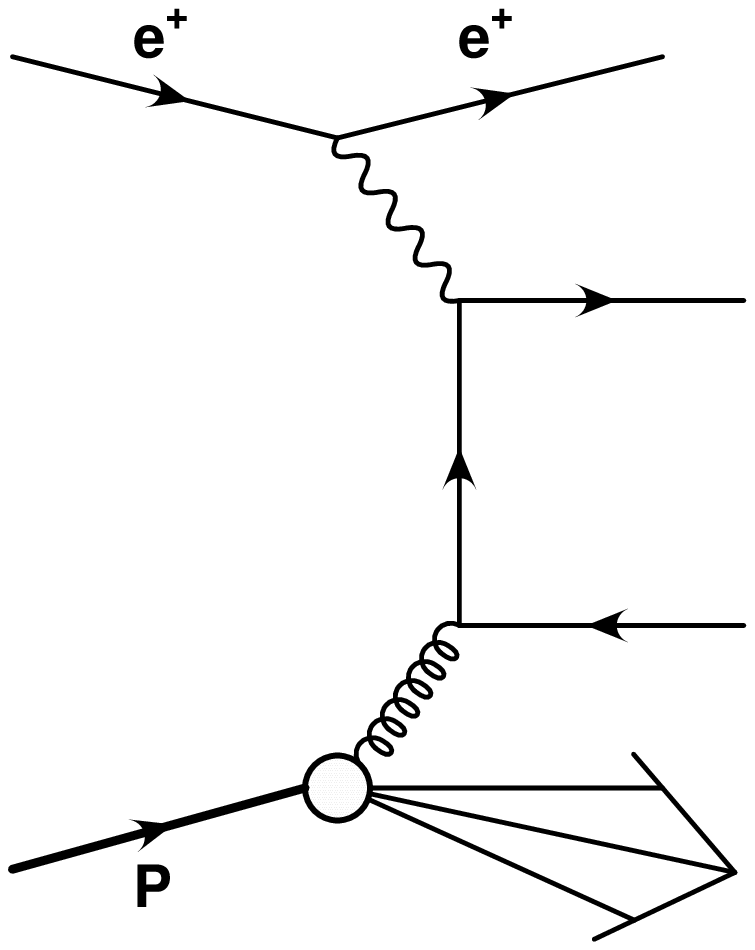,height=3.7cm}
\hspace{1cm}
\epsfig{file=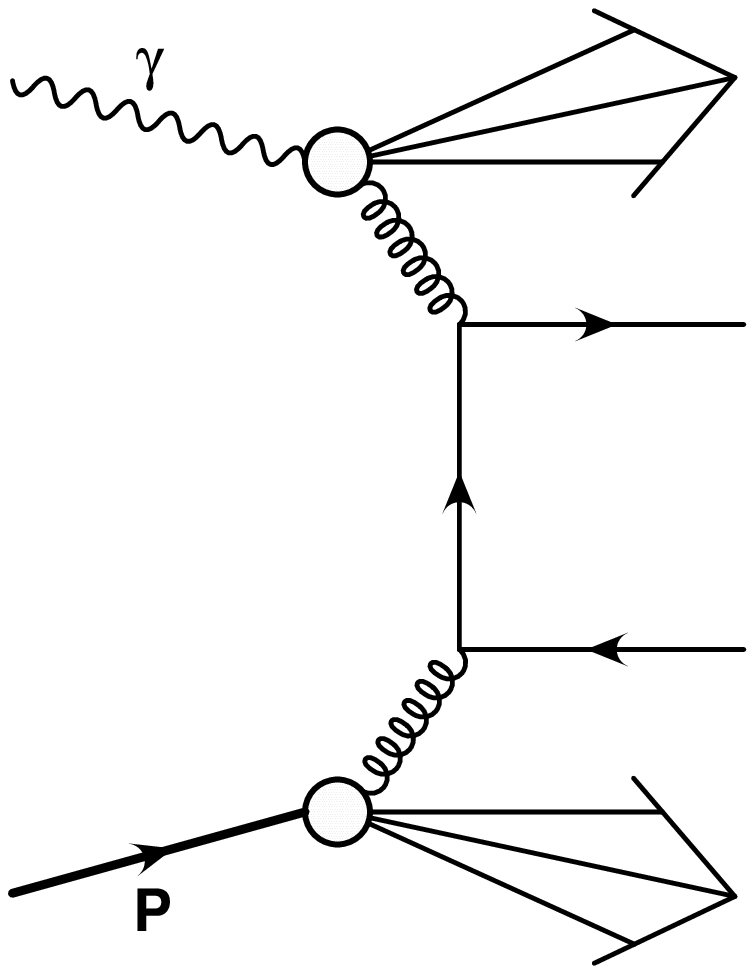,height=3.7cm}
\hspace{1cm}
\epsfig{file=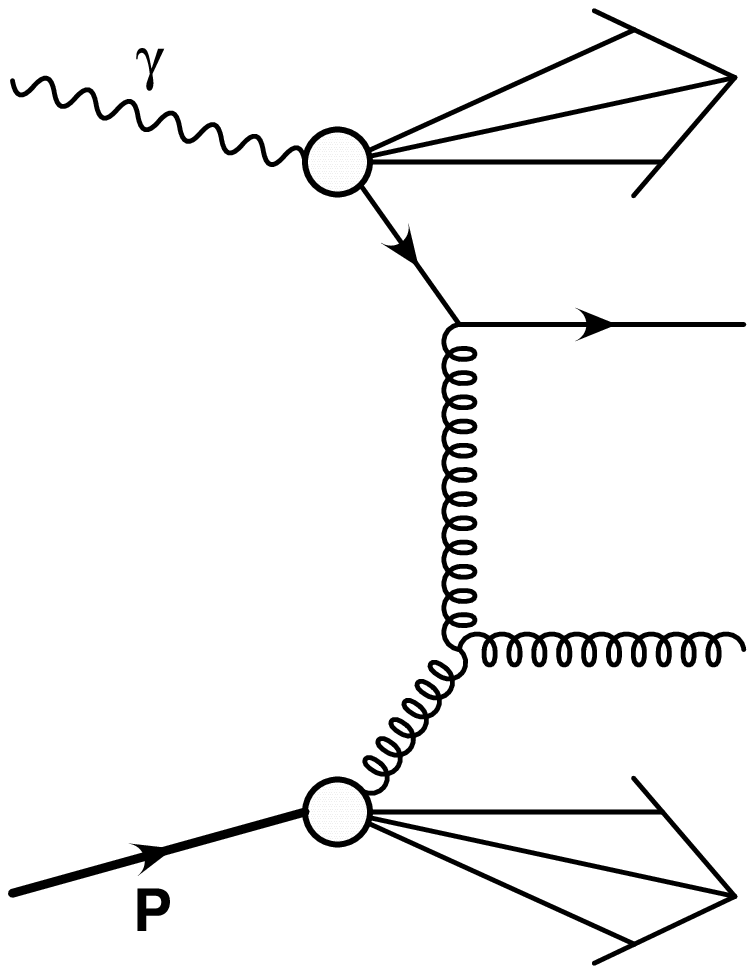,height=3.7cm}
\end{center}
\vspace{-0.4cm}
\hspace{2.15cm}(a) \hspace{3.575cm}(b) \hspace{3.55cm}(c)
\vspace{-0.1cm}
\caption[Examples of LO direct photon (a) and resolved photon (b) and
(c) processes.]
{Examples of LO direct photon (a) and resolved photon (b) and
(c) processes.}
\label{lo-feyn}
\end{figure}

The study of dijet photoproduction at HERA allows
one to perform tests of perturbative QCD (pQCD) and to place
constraints on the structure of the photon. To Leading Order (LO) two
types of processes contribute to jet photoproduction \cite{lo}
- direct and resolved photon processes (see Fig. \ref{lo-feyn}). In
direct photon processes, the photon interacts \emph{directly} in the
hard sub-process, whereas for resolved photon processes, the photon
\emph{resolves} into a source of partons, one of which participates in
the hard sub-process.

The cross-section for resolved photon processes can be written in
terms of the perturbatively calculable $2\rightarrow2$ scattering
cross-section, $d\sigma_{ab \rightarrow cd}$, as,

\begin{equation}
d\sigma_{\gamma p \rightarrow cd} = \sum_{ab} \int_{x_p}
\int_{x_\gamma} f_{p/b}(x_p, \mu_p^2) f_{\gamma/a}(x_\gamma,
\mu_\gamma^2) d\sigma_{ab \rightarrow cd}.
\label{pdf}
\end{equation}
The proton's parton density function, $f_{p/b}$, is
experimentally constrained allowing the extraction of the parton
density of the photon, $f_{\gamma/a}$, a currently poorly
known quantity.

Comparisons of data with pQCD calculations will be covered first
followed by analyses of the strucure of the photon.

\section{Tests of pQCD in Dijet Photoproduction}

The invariant mass of the dijet system, $M^{JJ}$, is sensitive to the
presence of new particles or resonances decaying into jets. The
scattering angle in the centre-of-mass frame of the dijet system,
$cos\theta^*$, reflects the underlying parton dynamics, thereby
providing a test of QCD. In direct photon processes, the dominant propagator
is that of a quark in the $s$, $t$ and $u$ channels but is a gluon
in the $t$ channel for resolved photon processes. The presence of the
different propagators in the two processes is reflected in the angular
dependence of the cross-section; the $spin-\frac{1}{2}$ quark yields a
$(1-|cos\theta^*|)^{-1}$ and the $spin-1$ gluon yields a
$(1-|cos\theta^*|)^{-2}$ dependence in the cross-section. This predicted
(by pQCD) behaviour of the cross-section dependence on direct and
resolved processes has been observed at HERA for $M^{JJ} > 23$
GeV \cite{costh}. Consequently any deviation from the pQCD predictions for the
cross-section for higher dijet invariant masses would also be an
indication of the presence of decays from new particles or resonances.

\begin{figure}[htp]
\vspace{-1.25cm}
\begin{center}
\epsfig{file=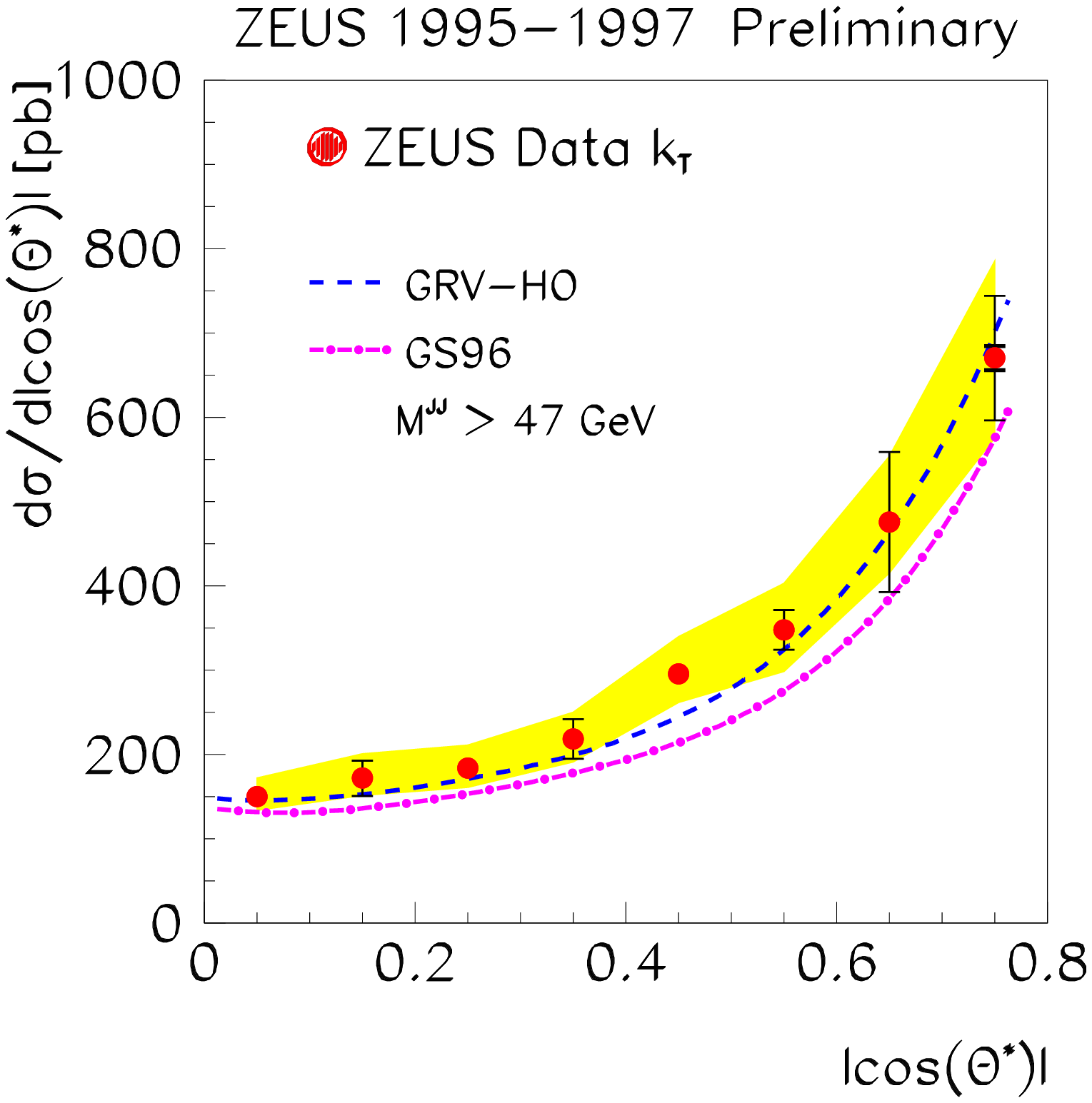,height=7cm}
\hspace{-1.5cm} \epsfig{file=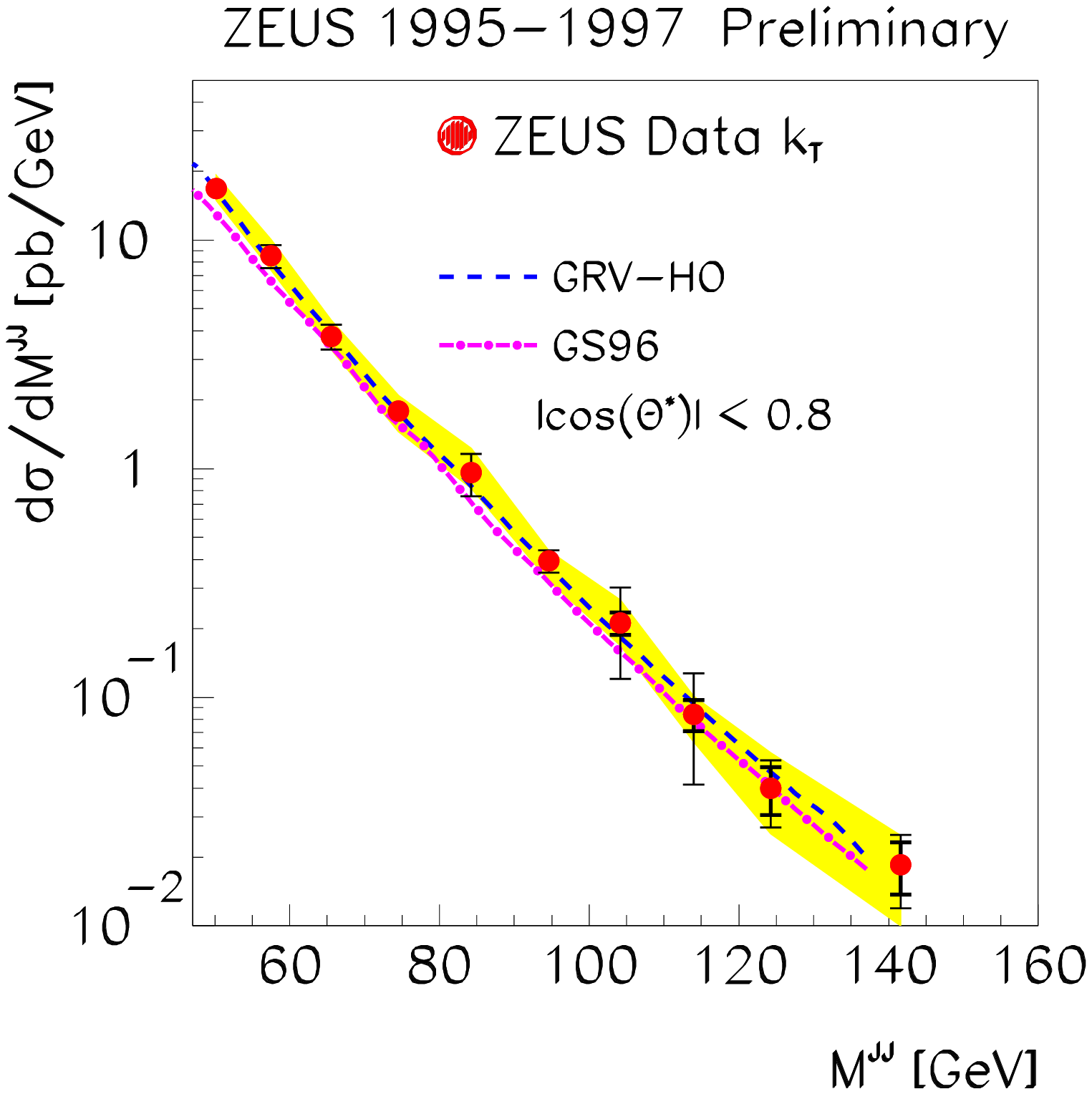,height=7cm}
\end{center}
\vspace{-0.95cm}
\caption[Differential cross-sections, $d\sigma/d|cos\theta^*|$ (left)
and $d\sigma/dM^{JJ}$ (right) compared to pQCD calculations. The ZEUS
data points show statistical errors (thick bars) and statistical plus
systematic errors (thin bars) with the error due to the energy scale
displayed as a band.]
{Differential cross-sections, $d\sigma/d|cos\theta^*|$ (left)
and $d\sigma/dM^{JJ}$ (right) compared to pQCD calculations. The ZEUS
data points show statistical errors (thick bars) and statistical plus
systematic errors (thin bars) with the error due to the energy scale
displayed as a band.}
\label{mjj}
\end{figure}

Photoproduction events were defined by requiring $Q^2 < 4 \ {\rm
GeV^2}$ (corresponding to a median $Q^2 \approx 10^{-3} \ {\rm GeV^2}$)
and the photon-proton centre-of-mass energy in the range, 134 $< W <$
277 GeV. Differential cross-sections as a function of $M^{JJ}$ and
$cos\theta^*$ for dijet masses above 47 GeV and $|cos\theta^*| < 0.8$
were measured. Jet finding was performed using the
$k_T$ \footnote{All measurements presented in this paper were
performed using the $k_T$ clustering algorithm} clustering algorithm
\cite{kt} requiring there to be at least two jets with $E_T^{jet}~>~14$~GeV and $-1~<~\eta^{jet}~<$~2. The measured cross-sections
$d\sigma/d|cos\theta^*|$  and $d\sigma/dM^{JJ}$ using 41.3 ${\rm
pb^{-1}}$  are shown in Fig. \ref{mjj} and compared to pQCD
calculations \cite{nlo-mjj}. 

The NLO calculations agree well in shape with the measured
distributions for both $d\sigma/d|cos\theta^*|$ and
$d\sigma/dM^{JJ}$. The predictions using the GRV-HO photon structure function
are closer in magnitude to the data compared to those
from GS96.

\section{Tests of pQCD in Multijet Photoproduction}

\begin{floatingfigure}[r]{6.2cm}
\centerline{\epsfig{file=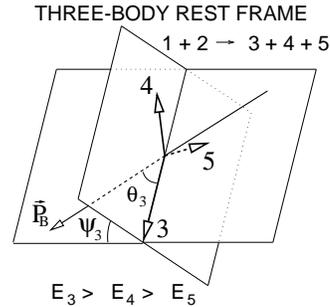,height=4.cm}}
\caption[Illustration of the angles $\theta_3$ and $\psi_3$ for a
particular three jet configuration.]
{Illustration of the angles $\theta_3$ and $\psi_3$ for a
particular three jet configuration.}
\label{3body}
\end{floatingfigure}
Multijet production provides a test of pQCD predictions beyond leading
order and additionally tests extensions to fixed order theories such
as parton shower models. The three jet system can be visualised in the
centre-of-mass frame by considering Fig. \ref{3body}. Of interest are
the angles $\theta_3$ and $\psi_3$. $\theta_3$ is the angle between
the highest energy jet and the beam direction and is analogous to the
scattering angle $\theta^*$ from the previous section. $\psi_3$ is the
angle between the plane containing the three jets and the plane
containing the highest energy jet and the beam direction.

\begin{floatingfigure}[l]{6.2cm}
\centerline{\epsfig{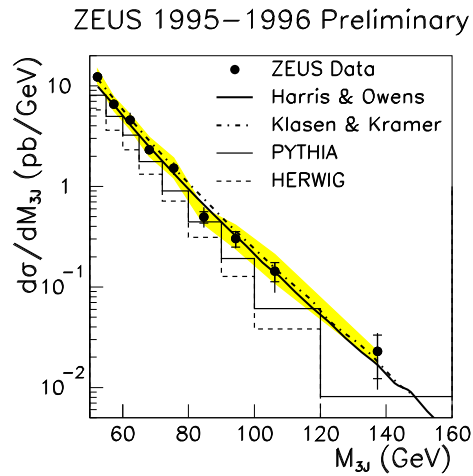}}
\vspace{-0.3cm}
\caption[Differential cross-section, $d\sigma/dM_{3J}$, compared to
pQCD calculations and Monte Carlo predictions. The ZEUS data points
show statistical errors (inner bars) and statistical plus systematic
errors (outer bars) with the error due to the energy scale displayed
as a band.]
{Differential cross-section, $d\sigma/dM_{3J}$, compared to pQCD
calculations and Monte Carlo predictions. The ZEUS data points show
statistical errors (inner bars) and statistical plus systematic errors
(outer bars) with the error due to the energy scale displayed as a
band.} 
\label{m3j}
\end{floatingfigure}
By requiring $Q^2<1 \ {\rm GeV^2}$ and a photon-proton energy range,
134~$<~W~<$~269~GeV, photoproduction events are selected. 
These events are then required to have at least two jets with
$E_T^{jet} >$ 6 GeV and a third with $E_T^{jet}~>~5$~GeV in a region
of pseudorapidity, $|\eta^{jet}| < 2.4$. These jet requirements
introduce a bias in the angular distributions by excluding those jets
produced close the beam-line. The criteria; $M_{3J}~>~50$~GeV,
$|cos\theta_3| <$ 0.8 and $2E_3/M_{3J} <$ 0.95, reduce this bias.

\begin{figure}[htp]
\centerline{\epsfig{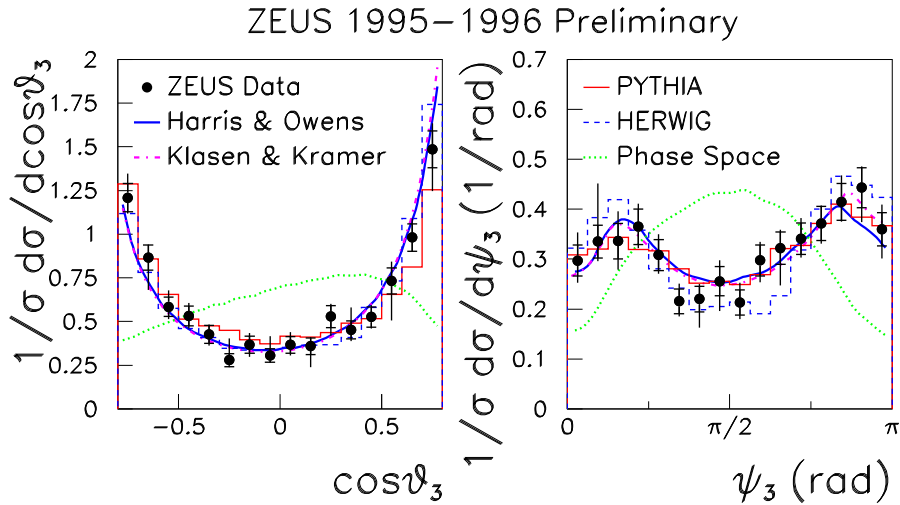}}
\caption[Distributions for $cos\theta_3$ (left) and $\psi_3$ (right). The statistical
and systematic errors are as described previously.]
{Distributions for $cos\theta_3$ (left) and $\psi_3$ (right). The statistical
and systematic errors are as described previously.}
\label{angles}
\end{figure}

The three jet invariant mass cross-section, $d\sigma/dM_{3J}$, is
shown in Fig. \ref{m3j} and compared with
$\mathcal{O}\mathit{(\alpha\alpha_s^2)}$ calculations from two pairs
of authors, Harris \& Owens \cite{hando} and Klasen \& Kramer
\cite{kandk}, showing good agreement with the data. The Monte Carlo
models HERWIG and PYTHIA describe the shape well but lie 20-40$\%$
below the data. In Fig. \ref{angles}, the angular distributions,
$cos\theta_3$  and $\psi_3$ are shown. These distributions show a
difference from those obtained from phase space displaying a
sensitivity to the QCD matrix elements. The $cos\theta_3$ distribution
is similar to that of a Rutherford scattering form and is well
predicted by the QCD calculations which account for the spin of the
propagator. On consideration of the distribution in $\psi_3$, one can
see a tendency for the three jet plane to lie near the plane
containing the beam and the highest energy jet as predicted by the
coherence property of QCD.

\section{High $E_T$ Dijet Cross-sections in Photoproduction}

By choosing events with suitably high $E_T$ dijets, one can study the
sensitivity of the cross-sections to the photon's parton distribution,
the proton's being well constrained from previous experiments. This
provides a measurement, in the range of the parton's momentum
fractions, $x_\gamma$, with a higher scale to those in $e^+e^-$
data. The current parameterisations of the parton density of the
photon in the high $x$ region have large differences due to
uncertainties in experimental measurements. In previous dijet
measurements \cite{bob}, a large excess of the measured cross-sections
for $E_T^{jet} >$ 6 GeV was seen over the NLO predictions for resolved
enriched samples. This was attributed to possible contributions from
non-perturbative effects such as multiparton interactions
\cite{mp}. This effect should be reduced for increasing
$E_T^{jet}$. To reconstruct $x_\gamma$ experimentally, we define the
observable:

\begin{eqnarray}
x_\gamma^{{\rm obs}} = \frac{\sum_{jets} E_T^{jet}
e^{-\eta^{jet}}}{2yE_e},
\label{xg}
\end{eqnarray}
where the sum runs over the two highest $E_T$ jets and $yE_e$ is the
initial photon energy. We can then define direct and resolved
processes by cutting on $x_\gamma^{{\rm obs}}$, such that direct
enriched requires $x_\gamma^{{\rm obs}} \geq$ 0.75.

\begin{floatingfigure}[r]{7cm}
\vspace{-0.75cm}
\centerline{\epsfig{file=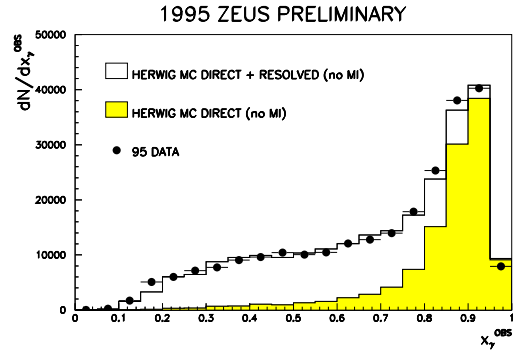,height=5.5cm}}
\vspace{-0.75cm}
\caption[The measured uncorrected $x_\gamma^{{\rm obs}}$ distribution
(dots) compared to HERWIG Monte Carlo predictions for direct only
(shaded histogram) and direct plus resolved (open histogram).]
{The measured uncorrected $x_\gamma^{{\rm obs}}$ distribution
(dots) compared to HERWIG Monte Carlo predictions for direct only
(shaded histogram) and direct plus resolved (open histogram).}
\label{xgamma}
\end{floatingfigure}
Cross-sections are measured in two kinematical regions of the photon
and proton centre-of-mass; the nominal region, 134 $< W <$ 277 GeV and
the region 212 $< W <$ 277 GeV, which enhances the low $x_\gamma^{{\rm
obs}}$ events. The photon virtuality was again required to be less
than 1 ${\rm GeV^2}$. Asymmetric requirements on the jet transverse
energy \cite{nlo-mjj} of $E_T^{jet1}>14$~GeV and $E_T^{jet2}>11$~GeV in the region $-1 < \eta^{jet} < 2$ were chosen. 

Fig.\ref{xgamma} shows the uncorrected $x_\gamma^{{\rm obs}}$
distribution compared to HERWIG Monte Carlo predictions. There is good
agreement in shape between the data and Monte Carlo without the
requirement of multiparton interactions. For lower transverse energy
jets ($E_T^{jet} >$ 6 GeV \cite{bob}), there was a large excess at low
$x_\gamma^{{\rm obs}}$ which is not seen here.

\begin{floatingfigure}[l]{7cm}
\vspace{-0.45cm}
\centerline{\epsfig{file=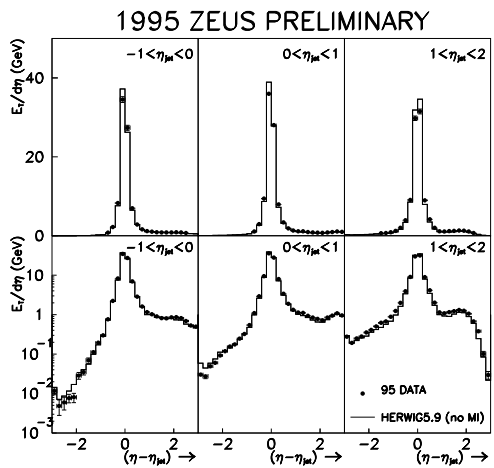,height=7cm}}
\vspace{-0.55cm}
\caption[Uncorrected transverse energy flow around the jets compared
to HERWIG Monte Carlo predictions for three pseudorapidity
regions. The flows are measured using the energy deposits in a band of
$\pm1$ unit from the jet centre in $\phi$ as a function of
$\Delta\eta = \eta^{cell}-\eta^{jet}$.]
{Uncorrected transverse energy flow around the jets compared
to HERWIG Monte Carlo predictions for three pseudorapidity
regions. The flows are measured using the energy deposits in a band of
$\pm1$ unit from the jet centre in $\phi$ as a function of
$\Delta\eta = \eta^{cell}-\eta^{jet}$.}
\label{profs}
\end{floatingfigure}
Evidence of the non-requirement of multiparton interactions to
describe the data can also be seen the distribution of transverse
energy flow around the jets. Consideration of Fig. \ref{profs} demonstrates
good agreement between data and Monte Carlo predictions. For high
values of $\eta$, the region of disagreement in lower $E_T$ dijet
studies \cite{bob}, the agreement remains good. From this observation,
coupled to the measurement of $x_\gamma^{{\rm obs}}$, we conclude that no
additional processes above the leading logarithm parton shower model
is necessary to describe the event shapes.

\begin{figure}[htp]
\vspace{-0.4cm}
\centerline{\epsfig{file=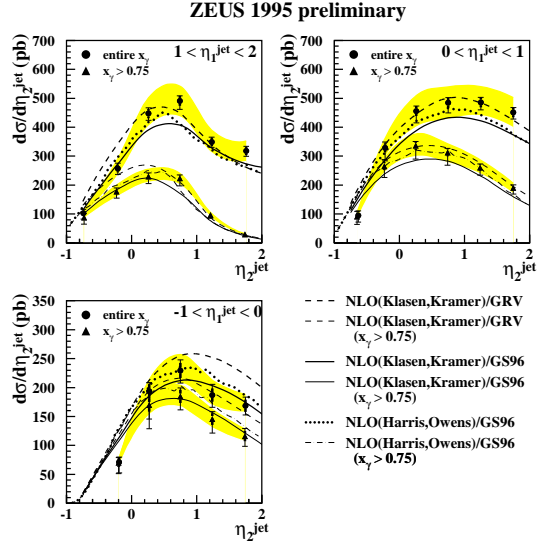,height=7.7cm}}
\vspace{-0.5cm}
\caption[Measured $d\sigma/d\eta_2^{jet}$ in 134 $< W <$ 277 GeV in
three regions of $\eta_1^{jet}$ for all $x_\gamma^{{\rm obs}}$
(circles) and for $x_\gamma^{{\rm obs}} >$ 0.75 (triangles) compared
to NLO calculations. The data has statistical errors (thick bars),
the sum of statistical and systematic errors (thin bars) and a band due to the
uncertainty in the energy scale.]
{Measured $d\sigma/d\eta_2^{jet}$ in 134 $< W <$ 277 GeV in
three regions of $\eta_1^{jet}$ for all $x_\gamma^{{\rm obs}}$
(circles) and for $x_\gamma^{{\rm obs}} \geq$ 0.75 (triangles) compared
to NLO calculations. The data has statistical errors (thick bars),
the sum of statistical and systematic errors (thin bars) and a band due to the
uncertainty in the energy scale.}
\label{ally}
\end{figure}

Differential cross-sections with respect to the pseudorapidity of the
second jet in bins of the pseudorapidity of the first jet are here
presented. For the entire region in $W$, $d\sigma/d\eta_2^{jet}$ is
shown for the whole region in $x_\gamma^{{\rm obs}}$ and for the
direct enriched ($x_\gamma^{{\rm obs}} \geq$ 0.75) region in
Fig. \ref{ally}. The data is compared to NLO calculations from Klasen
et al. \cite{kandk} and Harris et al. \cite{hetal} for the GS96 photon
structure function and the GRV-HO structure function for Klasen et
al.. Reasonable agreement between data and theory is seen in shape and
magnitude for both regions of $x_\gamma^{{\rm obs}}$. However,
differences of the order of the systematic errors are seen between the
two structure functions.

\begin{figure}[htp]
\vspace{-0.8cm}
\hspace{2.55cm} \epsfig{file=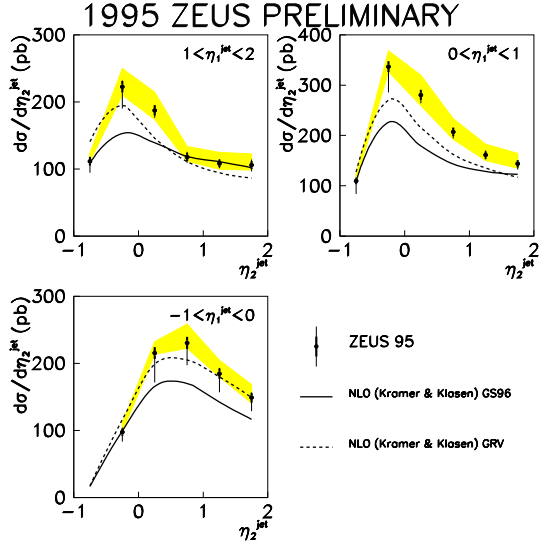,height=7.2cm}
\vspace{-0.3cm}
\caption[Measured $d\sigma/d\eta_2^{jet}$ in 212 $< W <$ 277 GeV in
three regions of $\eta_1^{jet}$ for all $x_\gamma^{{\rm obs}}$ compared
to NLO calculations. The data has statistical and systematic errors as
described previously.]
{Measured $d\sigma/d\eta_2^{jet}$ in 212 $< W <$ 277 GeV in
three regions of $\eta_1^{jet}$ for all $x_\gamma^{{\rm obs}}$ compared
to NLO calculations. The data has statistical and systematic errors as
described previously.}
\label{highy}
\end{figure}

Fig. \ref{highy} shows $d\sigma/d\eta_2^{jet}$ for the region of high
$W$ with all other cross-section definitons remaining the same. The
cross-sections for the whole range of $x_\gamma^{{\rm obs}}$ are shown
and compared to the calculations from Klasen et al. with the the two
previously mentioned structure functions. From these we see that this
high $W$ region leads to an increased sensitivity to the photon
structure function particularly in the $1<\eta_1^{jet}<2$ region. In
the central region of pseudorapidity the measured cross-sections lie
above the predictions.

\begin{figure}[htp]
\vspace{-0.15cm}
\centerline{\epsfig{file=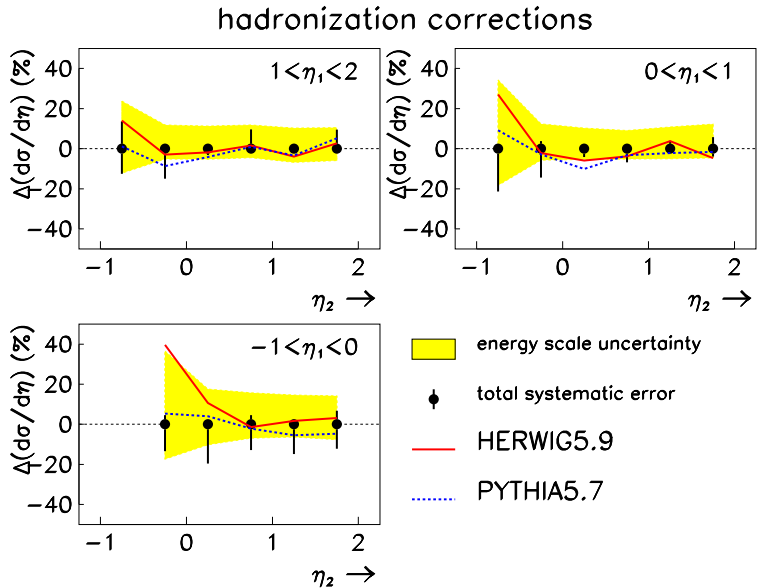,height=8.0cm}}
\vspace{-0.3cm}
\caption[Estimation of hadronization corrections,
$(d\sigma/d\eta_2^{jet})_{parton}/(d\sigma/d\eta_2^{jet})$-1. The
parton level cross-section, $(d\sigma/d\eta_2^{jet})_{parton}$ , is
calculated using the partons after the parton shower in the leading
logarithm Monte Carlos, HERWIG and PYTHIA. The errors are displayed as
previously defined.]
{Estimation of hadronization corrections,
$(d\sigma/d\eta_2^{jet})_{parton}/(d\sigma/d\eta_2^{jet})-1$. The
parton level cross-section, $(d\sigma/d\eta_2^{jet})_{parton}$ , is
calculated using the partons after the parton shower in the leading
logarithm Monte Carlos, HERWIG and PYTHIA. The errors are displayed as
previously defined.}
\label{hadcor}
\end{figure}

The uncertainty from possible differences between the jets from
final state particles and NLO partons due to higher order
contributions and hadronizations is \emph{partly} estimated using the
leading logarithm parton shower Monte Carlo, HERWIG and PYTHIA. The
estimator $(d\sigma/d\eta_2^{jet})_{parton}/(d\sigma/d\eta_2^{jet})-1$
for the narrower $W$ region, 212$<W<277$~GeV is shown in
Fig. \ref{hadcor} where the partons are those after the parton shower
process. The uncertainities are compared with the experimental
uncertainities for the measured $d\sigma/d\eta_2^{jet}$. For low
$\eta_2^{jet}$ the estimator from the HERWIG model is large (smaller
for PYTHIA) and comparable with the systematic uncertainties. The
theoretical uncertainty is smaller than the experimental error in the
other regions of $\eta^{jet}$ being similar for HERWIG and PYTHIA.

Another uncertainty comes from the ambiguity of the renormalisation
and factorisation scale, $\mu$, and is estimated by Harris et
al. \cite{hetal} to be of the order of 10\% by varying the scale;
$E_T/2$ to $2E_T$, for the cross-sections in Fig. \ref{ally}.
 
\section{Real and Virtual Photons in Dijet Production}

Whilst progress has recently been made in studying the parton
distribution functions (PDFs) for quasi-real photons, little
information exists for those of virtual photons with low statistics
data from the PLUTO collaboration and studies of the
photoproduction to deep inelastic transition region from the H1
collaboration \cite{pluto} being the only
published results. The expectation for the HERA data is that the
contribution to the cross-section from resolved photon compared to
direct photon processes should decrease with increasing photon
virtualities. Therefore measurements of the evolution of the resolved
photon component with $Q^2$ are expected to constrain the virtual
photon PDFs and test pQCD.

\begin{figure}[htp]
\vspace{-0.3cm}
\centerline{\epsfig{file=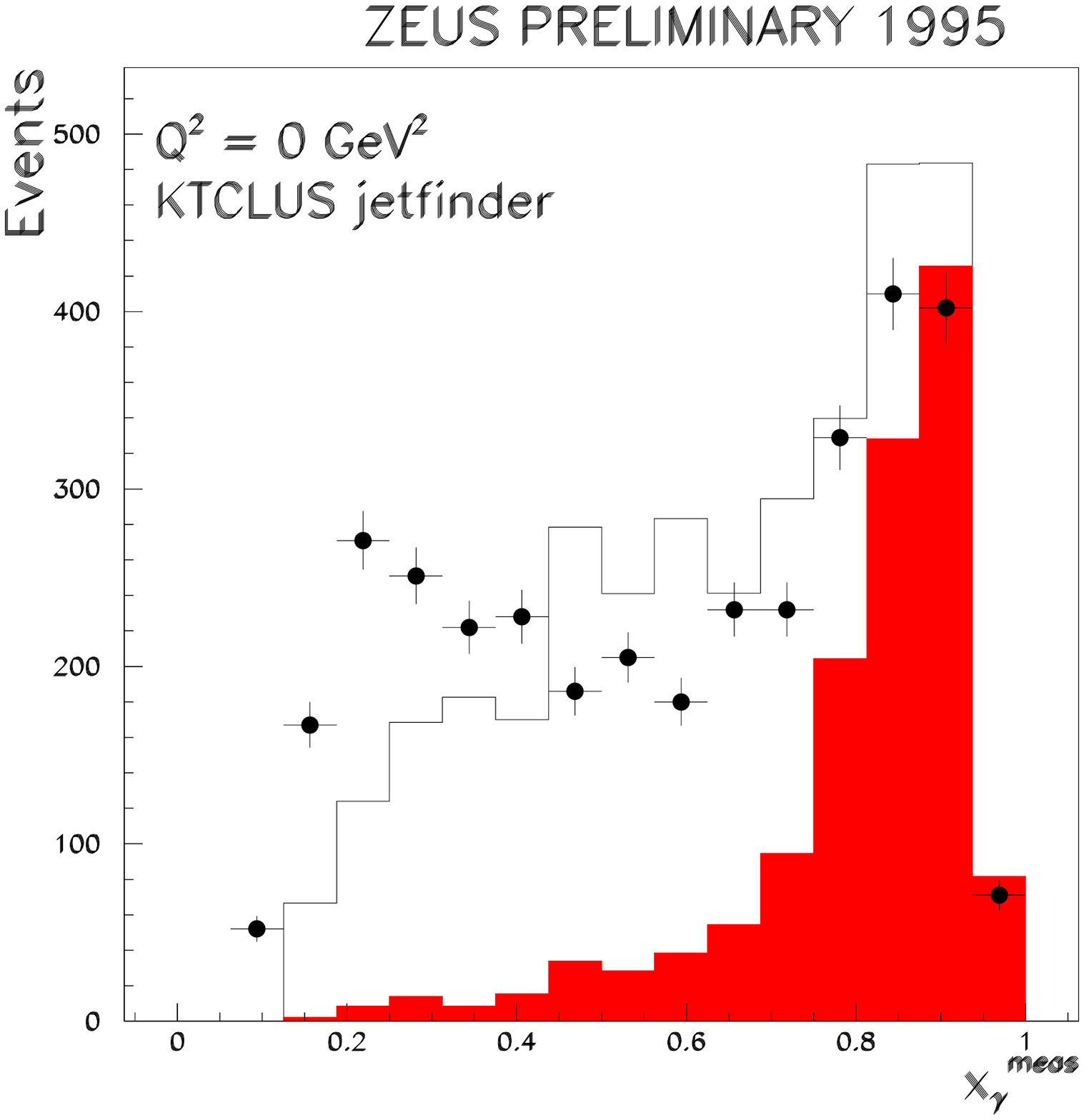,width=4.0cm,height=3.8cm} 
\epsfig{file=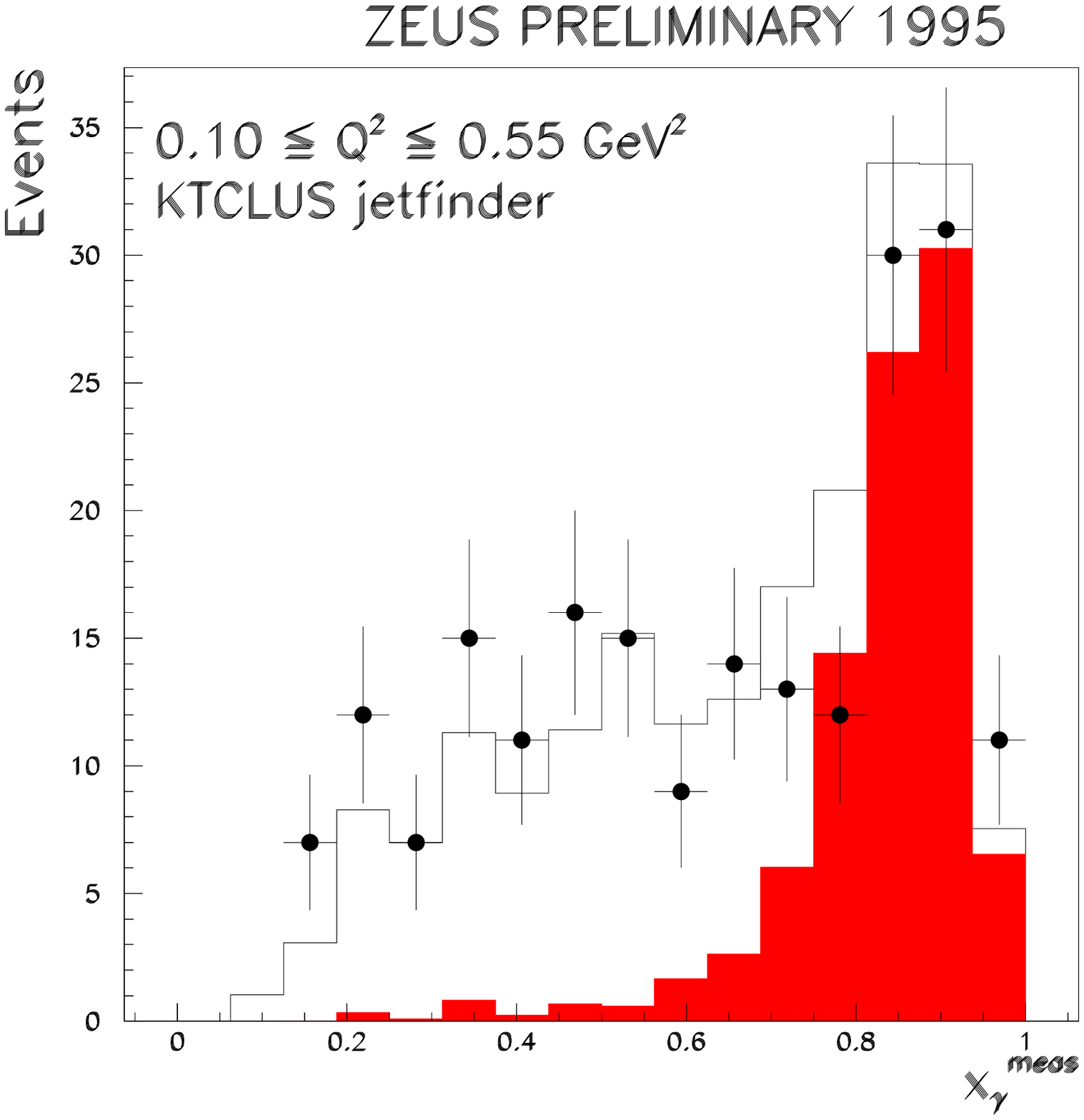,width=4.0cm,height=3.8cm}
\raisebox{-0.6ex}{\epsfig{file=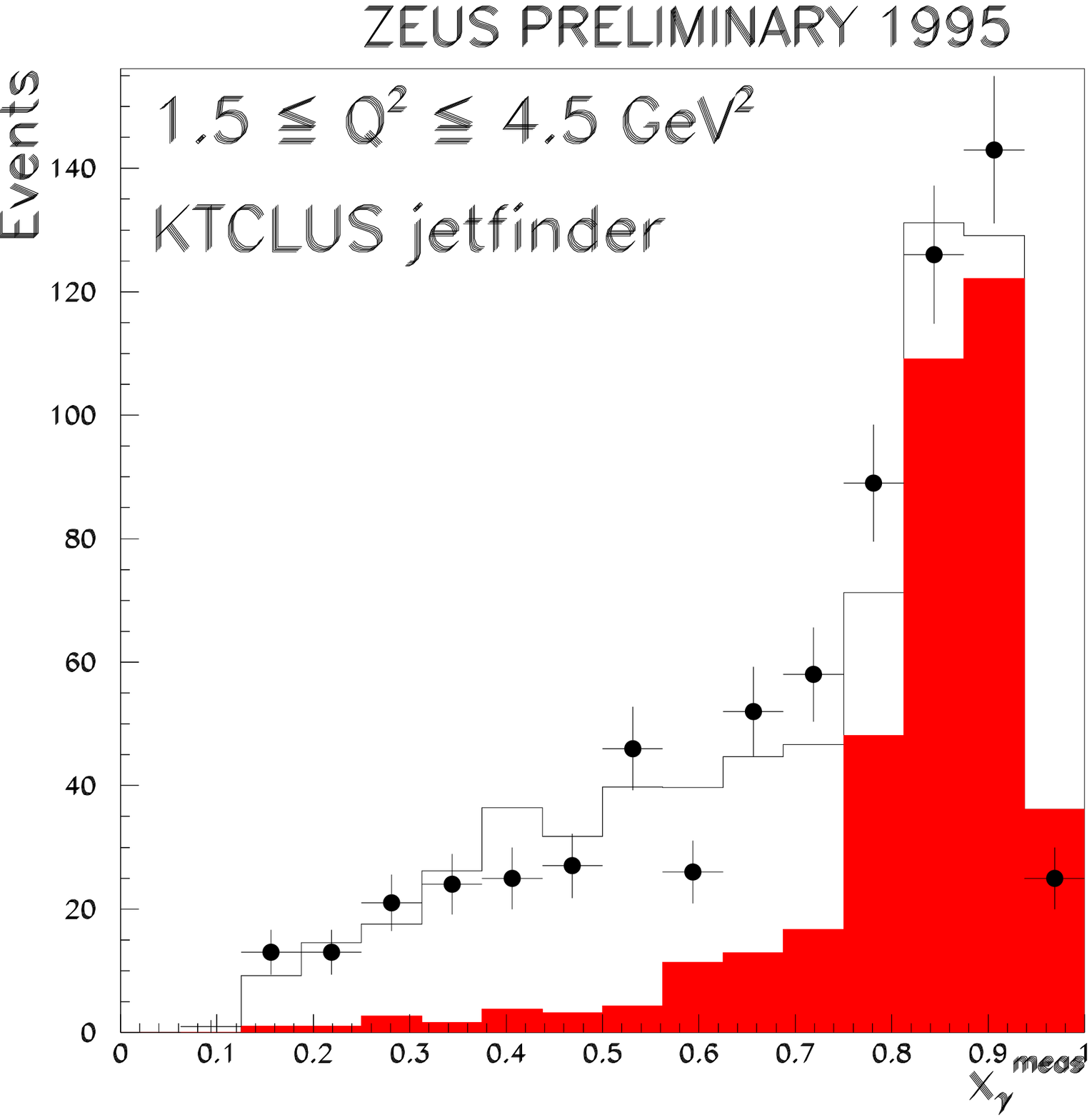,width=4.0cm,height=4.1cm,width=4.5cm}}}
\vspace{-0.3cm}
\caption[Uncorrected $x_\gamma^{{\rm meas}}$ in three regions of
photon virtuality compared to HERWIG Monte Carlo. Direct (shaded
histogram) and the sum of direct and resolved (open histogram) are
shown, where the ratio of the two was determined from a fit in each
bin of $Q^2$.]
{Uncorrected $x_\gamma^{{\rm meas}}$ in three regions of
photon virtuality compared to HERWIG Monte Carlo. Direct (shaded
histogram) and the sum of direct and resolved (open histogram) are
shown, where the ratio of the two was determined from a fit in each
bin of $Q^2$.}
\label{virtxg}
\end{figure}

\begin{floatingfigure}[r]{6.9cm}
\vspace{-0.7cm}
\centerline{~\epsfig{file=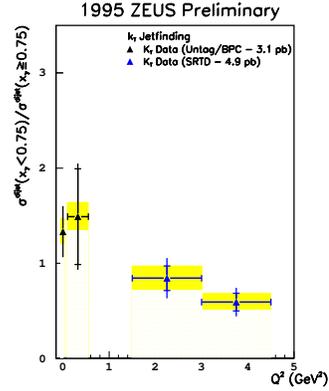,height=5.7cm}}
\vspace{-0.4cm}
\caption[The ratio, $\sigma(x_\gamma^{{\rm obs}} <
0.75)/\sigma(x_\gamma^{{\rm obs}} > 0.75$) as a function of photon
virtuality, $Q^2$. The points have statistical errors (inner bars) and
the sum of statistical and systematic (outer bars) and the band
represents the uncertainty due to the energy scale of the
calorimeter.]
{The ratio, $\sigma(x_\gamma^{{\rm obs}} < 0.75)/\sigma(x_\gamma^{{\rm obs}} > 0.75$) as a function of photon
virtuality, $Q^2$. The points have statistical errors (inner bars) and
the sum of statistical and systematic (outer bars) and the band
represents the uncertainty in the energy scale of the
calorimeter.}
\label{ratio}
\end{floatingfigure}
\vspace{-0.1cm}
As an estimator of $x_\gamma^{{\rm obs}}$, $x_\gamma^{{\rm meas}}$ was
defined which is analagous to Eqn. \ref{xg} but uses calorimeter
quantities; $y_{JB}$ as an estimator for $y$, $E_{T_{{\rm
meas}}}^{jet} > 5$ GeV and $-1.125 < \eta_{{\rm
meas}}^{jet} < 1.875$. Fig. \ref{virtxg} shows the $x_\gamma^{{\rm
meas}}$ distributions compared to HERWIG Monte Carlo predictions for
three bins of $Q^2$. The normalistaion of direct and resolved
processes were extracted from a two parameter fit to the measured
distribution, being different for each range in $Q^2$. The agreement
in shape for the virtual photon data is good but is poor for the
quasi-real photon data due to non perturbative effects as mentioned in
the previous section.

Using this data we can form the ratio of resolved enriched
($x_\gamma^{{\rm meas}} < 0.75$) to direct enriched ($x_\gamma^{{\rm
meas}}>0.75$) events and then compute the ratio of cross-sections,
$\sigma(x_\gamma^{{\rm obs}} < 0.75)/\sigma(x_\gamma^{{\rm obs}} >
0.75$), as a function of $Q^2$. This ratio defined for dijet
events, $E_T^{jet} >$ 6.5 GeV and -1.125 $< \eta <$ 1.875, with 0.2 $<
y <$ 0.55 is shown in Fig. \ref{ratio}. From this figure we can see a
general decrease in the ratio with increasing $Q^2$.

\section{Charm in Dijet Photoproduction}

Allied to the jets, a charm quark provides an additional hard scale
($m_c > \Lambda_{{\rm QCD}}$) yielding a more reliable perturbative
calculation. Two approaches for the calculation of charm are currently
available, the so-called \emph{massive} and \emph{massless}
schemes. The massive approach assumes only the three lightest quarks to be active flavours in the proton and photon. In the massless
approach, the charm quark is treated as an additional active
flavour. Massless calculations therefore predict, for a given
factorisation scale, a larger resolved component compared to massive
calculations. A measurement of $x_\gamma^{{\rm obs}}$ will provide a
method for the possibilty of distinguishing the two schemes and
probing the question of charm in the photon.

\begin{floatingfigure}[r]{6.8cm}
\vspace{-0.5cm}
\centerline{\epsfig{file=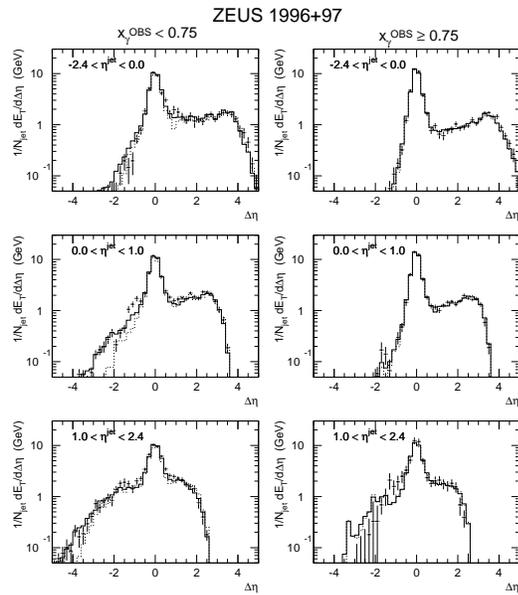,height=7.8cm}}
\vspace{-0.3cm}
\caption[Uncorrected transverse energy flow with respect to the jet
axis compared to HERWIG Monte Carlo (full histogram) and LO-direct
photon Monte Carlo only (dotted histogram). The distributions are
separated into low and high $x_\gamma^{{\rm obs}}$ for three bins in
$\eta^{jet}$.]
{Uncorrected transverse energy flow with respect to the jet
axis compared to HERWIG Monte Carlo (full histogram) and LO-direct
photon Monte Carlo only (dotted histogram). The distributions are
separated into low and high $x_\gamma^{{\rm obs}}$ for three bins in
$\eta^{jet}$.}
\label{charmprof}
\end{floatingfigure}
The charm quark itself cannot be directly tagged, consequently events
are selected containing reconstructed $D^{*\pm}$ mesons with
$p_T(D^*) > 3$ GeV/c and
$|\eta(D^*)| < 1.5$ in dijet photoproduction events. Before measuring
a cross-section in $x_\gamma^{{\rm obs}}$, we first consider the
energy flow about the jet axis, Fig. \ref{charmprof}. The jet
profiles, calculated in the same way as described in Fig. \ref{profs},
are split into low and high $x_\gamma^{{\rm obs}}$ for three bins of
$\eta^{jet}$ for reconstructed jets; $E_T^{CALjet} >$ 4 GeV. The HERWIG
Monte Carlo \emph{without} multiparton interactions provides a good
description of the data even in the forward region where there was a
discrepancy for inclusive dijets of this energy \cite{bob}. One can
also see that the direct photon only Monte Carlo cannot describe the
data at low values of $\Delta \eta$ as can be seen in the
$x_\gamma^{{\rm obs}} <$ 0.75, $0 < \eta^{jet} < 1$ bin. This excess
energy flow in the rear direction is consistent with there being a
photon remnant. 

Fig. \ref{charmxg} shows $d\sigma/dx_\gamma^{{\rm obs}}$ in the range
$Q^2 < 1 \ {\rm GeV^2}$, 130 $< W <$ 280 GeV for jets with $|\eta| <
2.4$, $E_T^{jet1,2} > 7,6$ GeV and at least one $D^*$ meson,
satisfying the criteria previously mentioned. 

\pagebreak

Fig. \ref{charmxg}(a) shows the HERWIG Monte Carlo (normalised to the
data) agreeing in 
\begin{floatingfigure}[r]{7.cm}
\epsfig{file=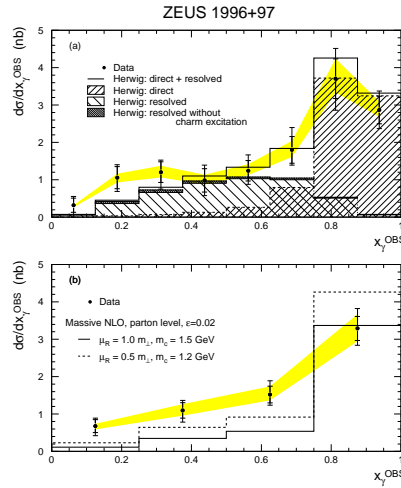,height=6.4cm}
\vspace{-0.25cm}
\caption[Differential cross-section, $d\sigma/dx_\gamma^{{\rm obs}}$
for dijets with an associated $D^*$ meson compared to HERWIG Monte
Carlo predictions (a) and a massive NLO calculation (b). The inner
(outer) error bars represent the statistical (statistical plus
systematic) errors and the band represents the uncertainty in the
energy scale.]
{Differential cross-section, $d\sigma/dx_\gamma^{{\rm obs}}$
for dijets with an associated $D^*$ meson compared to HERWIG Monte
Carlo predictions (a) and a massive NLO calculation (b). The inner
(outer) error bars represent the statistical (statistical plus
systematic) errors and the band represents the uncertainty in the
energy scale.}
\label{charmxg}
\end{floatingfigure}
\hspace{-0.58cm} shape with the measured cross-section.  There is
a peak at high $x_\gamma^{{\rm obs}}$ \ consistent with LO-direct
photon processes. There also exists a significant cross-section at low
$x_\gamma^{{\rm obs}}$ which cannot be described by LO-direct only and
needs some component of LO-resolved photon processes. At the given
scale the LO-resolved component is dominated by charm excitation
processes of the form, Fig. \ref{lo-feyn}(c). The required LO-resolved
contribution is 45$\pm$5(\emph{stat.})\% (compared to the HERWIG
prediction of 37\%) of which 93\% is from charm excitation processes.

Fig. \ref{charmxg}(b) shows a comparison of the data with an NLO
massive calculation \cite{frix}. This does not describe the low
$x_\gamma^{{\rm obs}}$ measured cross-section whilst describing the
high $x_\gamma^{{\rm obs}}$ region. Upon choosing extremes of the
scale $\mu_R$ and the charm mass $m_c$, the prediction still fails to
describe the data.

\vspace{-0.35cm}

\section{Summary and Conclusions}

Sufficient data has now being collected by ZEUS to allow the comparison
of pQCD with precise measurements; these comparisons show good agreement
for dijet and multijet dynamics. With the increased precision the
data can now differentiate between different photon structure function
parameterizations and can be used to place constraints on its form.

\end{document}